\documentclass[
nofootinbib,
amsmath,amssymb,
prd,
twocolumn
]{revtex4-1}

\usepackage[utf8]{inputenc}
\usepackage{epsf, float, graphicx,dcolumn}
\usepackage{}
\usepackage{hyperref}
\usepackage{multirow}
\usepackage{booktabs}
\usepackage{siunitx}
\usepackage{soul}
\usepackage{xcolor}
\usepackage{blindtext}

\hypersetup{colorlinks=true,linkcolor=blue,citecolor=blue,filecolor=blue,urlcolor=blue}

\usepackage{scalerel}
\usepackage{tikz}
\usetikzlibrary{svg.path}

\definecolor{orcidlogocol}{HTML}{A6CE39}
\tikzset{
  orcidlogo/.pic={
    \fill[orcidlogocol] svg{M256,128c0,70.7-57.3,128-128,128C57.3,256,0,198.7,0,128C0,57.3,57.3,0,128,0C198.7,0,256,57.3,256,128z};
    \fill[white] svg{M86.3,186.2H70.9V79.1h15.4v48.4V186.2z}
                 svg{M108.9,79.1h41.6c39.6,0,57,28.3,57,53.6c0,27.5-21.5,53.6-56.8,53.6h-41.8V79.1z M124.3,172.4h24.5c34.9,0,42.9-26.5,42.9-39.7c0-21.5-13.7-39.7-43.7-39.7h-23.7V172.4z}
                 svg{M88.7,56.8c0,5.5-4.5,10.1-10.1,10.1c-5.6,0-10.1-4.6-10.1-10.1c0-5.6,4.5-10.1,10.1-10.1C84.2,46.7,88.7,51.3,88.7,56.8z};
  }
}
\newcommand\orcidicon[1]{\href{https://orcid.org/#1}{\mbox{\scalerel*{
\begin{tikzpicture}[yscale=-1,transform shape]
\pic{orcidlogo};
\end{tikzpicture}
}{|}}}}

\newcommand{\github}{\href{https://github.com/James11222/CosmoMMF}{\textsc{cosmommf github}}}

\begin{document}

\title{Effects of Baryonic Feedback on the Cosmic Web}

\author{James Sunseri \orcidicon{0000-0003-4274-2662}}
\email{jamessunseri@berkeley.edu}
\affiliation{$^{1}$Department of Astronomy, University of California, Berkeley, CA 94720-3411, USA}
\affiliation{$^{2}$Department of Physics, University of California, Berkeley, CA 94720-7300, USA}

\author{Zack Li \orcidicon{0000-0002-0309-9750}}
\affiliation{Canadian Institute for Theoretical Astrophysics, University of Toronto, Toronto, ON, Canada M5S 3H8}

\author{Jia Liu \orcidicon{0000-0001-8219-1995}}
\affiliation{Kavli IPMU (WPI), UTIAS, The University of Tokyo, Kashiwa, Chiba 277-8583, Japan}

\date{\today}

\begin{abstract}
We study the effect of baryons on the cosmic web --- halos, filaments, walls, and voids.  To do so, we apply a modified version of NEXUS, a cosmic web morphological analysis algorithm, to the IllustrisTNG simulations. We find that halos lose more than 10\% of their mass due to baryons, mostly to filaments and a small portion to walls and voids. However, the mass transfer does not significantly shift the boundaries of structures, leaving the volume fractions of the cosmic structures largely unaffected. We
quantify the effects of baryonic feedback on the power spectrum and the probability density function (PDF) of the density field for individual cosmic structures. For the power spectrum, most suppression due to feedback can be accounted for by including $M\ge10^{12}~M_\odot/h$ halos, without considering other cosmic structures. However, when examining the PDF of the density field, we find nearly 100\% suppression of the emptiest regions and 10\%-level effects (boost or suppression) in the remaining regions of filaments, walls, and voids. Our results indicate the importance of modeling the effects of baryons in the whole cosmic web, not just halos, for cosmological analysis beyond two-point statistics or field-based inferences. Our code is available through \github. 
\end{abstract}

\maketitle

\section{Introduction}
\label{sec:intro}

The impact of baryonic feedback on the distribution of matter is poorly understood, and remains a major systematic for ongoing and upcoming surveys. Modern cosmological probes rely primarily on accurate models of the density field, using a combination of analytic theories and numerical simulations. However, baryonic processes such as active galactic nuclei (AGN) and supernova feedback redistribute matter in the universe, resulting in uncertainties in small-scale matter clustering with respect to gravity-only predictions (see a recent review by~\cite{chisari2019}). Their signature resembles those of key cosmological parameters related to dark energy, dark matter, modified gravity, and neutrino mass. Therefore, baryonic feedback must be carefully studied to achieve unbiased results with upcoming surveys such as the
Rubin Observatory LSST\footnote{\url{https://www.lsst.org}},
Euclid\footnote{\url{https://www.euclid-ec.org}}, 
SPHEREx\footnote{\url{https://www.jpl.nasa.gov/missions/spherex}}, and Roman Space Telescope\footnote{\url{https://wfirst.gsfc.nasa.gov/index.html}}. 

Cosmic structures beyond halos, such as filaments, walls, and voids, have been studied intensively in recent years~\cite{cweb_Bond1996,cweb_Shandarin2004,cweb_Bos2012,cweb_Ricciardelli2014,cweb_Park2009,cweb_Gheller2015,cweb_Martizzi2019}. They have been shown to contain rich cosmological information~\cite{cweb_Davies2018,cweb_Kreisch2019,cweb_Hamaus_2020,cweb_Coulton2020,cweb_bayer2021detecting}. To quantify the effect of baryons on these structures, in this work we apply a modified version of the cosmic web morphological analysis algorithm NEXUS~\cite{NEXUS_Cautun_2012,Cosmic_Web_Cautun} on the hydrodynamic simulation IllustrisTNG~\cite{Pillepich2018_TNG, Naiman2018_TNG, Nelson2018_TNG, Springel2018_TNG, Marinacci2018_TNG}. We study the effect of baryons individually and jointly on halos, filaments, walls, and voids by comparing these structures identified in the hydrodynamic simulation to those in the dark matter (DM) only simulation. We quantify our results as changes in: (1) the mass fractions and (2) the volume fractions of individual structures, (3) the matter power spectrum, and (4) the probability density function (PDF) of the density field. 

Our work scrutinizes the common assumption that baryons only affect the matter distribution within halos, made by halo-based models. For example, at the summary statistics-level, baryonic feedback on the matter power spectrum is quantified by changing the halo profiles~\cite{Rudd2008,analytic_Semboloni2011,analytic_Mohammed2014,Velliscig2014,analytic_Mead2015,analytic_Copeland2018,Mead2021}, based on the halo model~\cite{analytic_Peacock2000,analytic_Seljak2000}; 
at the field-level, similar philosophy is adopted by models that operate on three-dimensional particle data, 
where the particles within halos are displaced in dark-matter only simulations to mimic hydrodynamic effects~\cite{baryonification_Schneider2015, baryonification_Schneider2019,Arico2021,Lu2022,Lee2022}. These halo-based models include a handful of parameters that can be calibrated or marginalized over, using hydrodynamic simulations or observational data. Meanwhile, there exist other methods to quantify the impacts of baryons which do not rely on halos, e.g. the principal component analysis method~\cite{ML_Eifler2015, ML_Mohammed2018, Huang2019}, the Enthalpy Gradient Descent method~\cite{Dai2018}, and the Lagrangian deep learning method~\cite{Dai2021}. Our work aims to identify the regimes where the halo-only assumptions may fail and therefore more general models should be adopted. 

The paper is structured as follows:  we outline the simulations and methodologies in our analysis in Section \ref{sec:methods}, discuss our results in Section \ref{sec:results}, and finally conclude in Section \ref{sec:conclusion}.

\section{Methodology}
\label{sec:methods}
In this section, we describe the methodology for our analysis. We start this section with our choice of simulations used for the analysis in Section \ref{subsec:simulations}. We then discuss our methods for tagging cosmic structures with a modified NEXUS algorithm in Section \ref{subsec: nexus}. Lastly, we discuss how we measured the power spectrum and PDF in Section \ref{subsec:statistics}.

\subsection{Simulations}
\label{subsec:simulations}
We use the IllustrisTNG simulations~\cite{Pillepich2018_TNG, Naiman2018_TNG, Nelson2018_TNG, Springel2018_TNG, Marinacci2018_TNG}, a set of hydrodynamic simulations in cosmological volumes carried out by the moving-mesh code \textsc{arepo} \cite{sim_Springel2010, sim_Springel2022}. We use the highest resolution runs of the largest box size, TNG-300 and TNG-300-Dark simulations, corresponding to the hydrodynamic and dark matter-only runs, respectively. The hydrodynamic simulation includes sub-grid models that describe star formation, stellar evolution, chemical enrichment, primordial and metal-line cooling of the gas, stellar feedback with galactic outflows, black hole formation, growth, and multimode feedback. 
Both simulations share the same initial conditions and have periodic boundary conditions with side length $L = 205 h^{-1} \mathrm{Mpc}$. The cosmological parameters are set as $\Omega_{\rm m} = 0.3089$, $\Omega_{\rm b} = 0.0486$, $\Omega_{\lambda} = 0.6911$, and $h = 0.6774$. 

To apply the cosmic web classification on the simulations, we create density grids from individual simulations using \textsc{nbodykit} \cite{nbodykit_Hand_2018}. We generate three-dimensional (3D) regular grids of $1024^3$ in size. While we only need to consider dark matter particles for TNG-300-Dark, for TNG-300, we take into consideration all forms of mass, including dark matter, gas, stars, and black hole particles. We focus our analysis on $z=0$ snapshots as they have the longest integrated time of baryonic effects and the most non-linear structures. 

\begin{figure*}[ht]
    \centering
    \includegraphics[width=2\columnwidth]{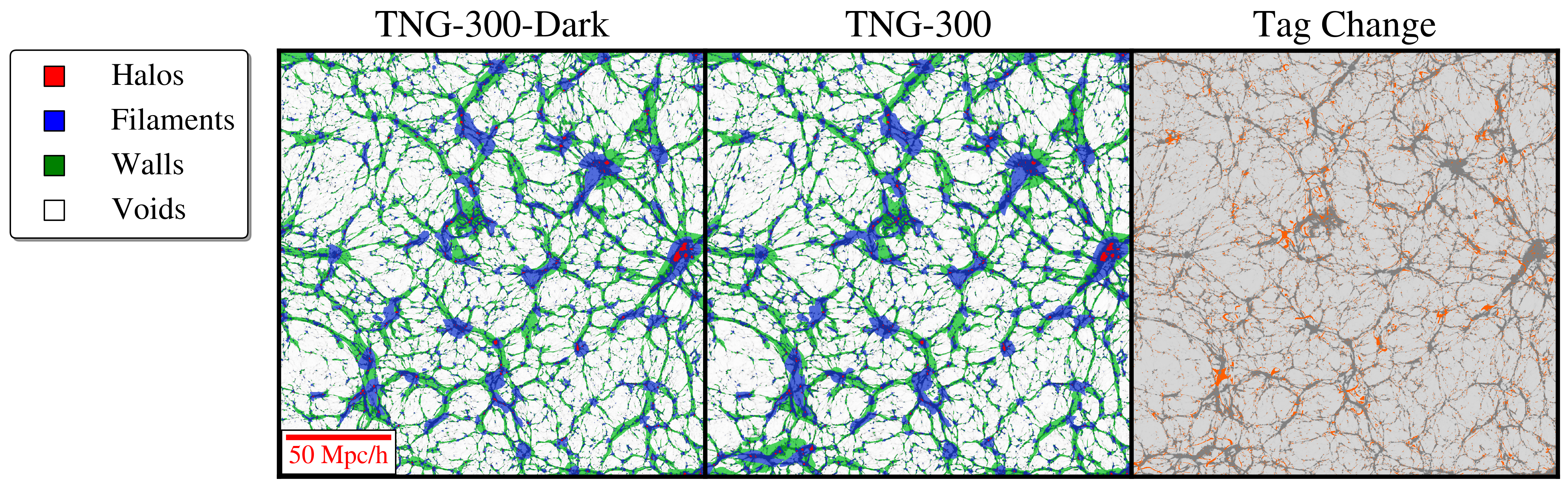}
    \caption{Cosmic structures (red: halos, blue: filaments, green: walls, white: voids) in a $0.2 \; h^{-1} \rm Mpc$ slice of the TNG-300-Dark (left) and TNG-300 (middle) simulations. The right panel shows changes in identification between the two simulations denoted by orange, with the density field shown as grey background for reference.}
    \label{fig:tagging}
\end{figure*}

\subsection{Cosmic Structure Identification}
\label{subsec: nexus}
We adopt the NEXUS algorithm, a multiscale morphological analysis tool developed by~\cite{NEXUS_Cautun_2012}, to identify cosmic structures within the simulations. However, we replace the density Hessian-based halo identification scheme in NEXUS with a halo catalog-based tagging method using the halo catalogs provided by the IllustrisTNG team. Our code is implemented in the \texttt{Julia} language and available through \github.\footnote{\href{https://github.com/James11222/CosmoMMF}{https://github.com/James11222/CosmoMMF}} We refer readers to the original NEXUS paper for more details.  

Here, we briefly introduce the procedure to identify cosmic structures in four main steps:
\begin{enumerate}
    \item \textbf{Halo Identification:} First, we identify the ``halo'' voxels using the halo catalog provided by the IllustrisTNG team. The halos are identified with the Friends-of-Friends (FoF) algorithm \citep{FOF_Press_and_Davis_1982, FOF_Huchra_and_Geller_1982, FOF_Davis_1985} with a linking length of $b = 0.2$. The FoF algorithm is run on the dark matter particles only, then the gas, stars, black hole particles are attached to the same halo as their nearest dark matter particle. We also investigate the effect of changing minimal halo mass, with $M_{\rm min}$=$[10^{11}, 10^{12}, 10^{13}]~M_\odot/h$.
    \item \textbf{Density Field Smoothing:} Next, we obtain the smoothed log-density field $f_{R}(\mathbf{x})$ with a Gaussian filter of radius $R$, in preparation for identifying filaments and walls. It is necessary to smooth the raw 3D density field first so that the input field for the NEXUS algorithm is differentiable. We use a smoothed log-density field instead of a smoothed density field because the former returns sharper structures in comparison. \footnote{In the NEXUS paper, when the log-density field is used, the algorithm is called NEXUS+ (see section 3.1.1 of ~\cite{NEXUS_Cautun_2012}). We adopt the NEXUS+ method throughout this paper. However, we do not make a distinction here for simplicity, as NEXUS and NEXUS+ are largely similar in implementation.} We apply 10 smoothing scales $R$ with logarithmic increments of $\sqrt{2}$. $R / R_0 \in \{1, \sqrt{2}, 4, ..., 16, 16\sqrt{2}\}$ where $R_0 = 0.2 h^{-1} \mathrm{Mpc}$ is the physical scale of one voxel. Multiple smoothing scales are necessary to cover the wide range of characteristic scales of cosmic structures. 
    \item \textbf{Hessian Computation:} We then compute the Hessian matrix of the smoothed field and the eigenvalues of the Hessian for each scale $R$, 
\begin{align}
    \mathbf{H}_{ij,R}(\mathbf{x}) =& R^2 \frac{\partial^2 f_{R}(\mathbf{x}) }{\partial x_i \partial x_j}\\
    \mathrm{det}(\mathbf{H}_{ij} - \lambda_{a}\mathbf{I}) =& 0, {\rm \; with\;} \lambda_1 \leq \lambda_2 \leq \lambda_3 
\end{align}
where $\mathbf{H}_{ij}$ is the Hessian of the log-smoothed field $f_{R}(\mathbf{x})$,  $\lambda_{a}$ are the eigenvalues with $a \in \{1,2,3\}$, and $\mathbf{I}$ is the identify matrix. $R$ represents the smoothing scales used in the density field smoothing, and its square is used to normalize the Hessian across different smoothing scales.
    \item \textbf{Structure Classification:} Based on the computed eigenvalues, we can then identify ``filament'' and ``wall'' voxels. Filaments are restricted to voxels with $\lambda_1<0, \lambda_2<0$, while walls  with $\lambda_1<0$. From this definition, it is apparent that any filament voxels would also satisfy as walls. Therefore, we need to define the relative strength of ``filament'' versus ``wall'' signatures:
    \begin{align}
    \mathcal{S}^f_R(\mathbf{x}) = &  \frac{\lambda_2^2}{|\lambda_1|} \left(1 - \left|\frac{\lambda_3}{\lambda_1}\right|\right),\\
    \mathcal{S}^w_R(\mathbf{x}) =&  |\lambda_1| \left(1 - \left|\frac{\lambda_2}{\lambda_1}\right| \right) \left(1 - \left|\frac{\lambda_3}{\lambda_1}\right| \right).
    \end{align}
    The $R$-dependent signatures are computed for the 10 smoothing scales, and the final signature strength is taken as the max of the 10 values,
\begin{equation}
    \mathcal{S} (\mathbf{x}) = \max\limits_{\textrm{n=1,2,..10}} \mathcal{S}_{R_n} (\mathbf{x}).
\end{equation}
The threshold $\mathcal{S}_{\rm th}$ for a voxel to be considered as filament or wall corresponds to the peak of the mass change with respect to the corresponding signature,
\begin{equation}
    \Delta M^2 = \left| \frac{d M^2}{d \log \mathcal{S}} \right|,
\end{equation}
where $M$=$M_f$($M_w$) is the mass in filaments (walls) for a given signature $\mathcal{S}^f$($\mathcal{S}^w$). We tag any voxels with $\mathcal{S}^f>\mathcal{S}^f_{\rm th}$  as ``filament''. We repeat the same process on the remaining voxels (non-cluster and non-filament) so that voxels with wall signatures greater than the wall threshold $\mathcal{S}^w>\mathcal{S}^w_{\rm th}$ are tagged as ``wall'' voxels. \footnote{Detailed discussions on optimal filament and wall separation can be found in Appendix A of the NEXUS paper~\cite{NEXUS_Cautun_2012}.} Finally, any voxels left unidentified are tagged as ``void''. 
\end{enumerate}

We show an example of identified cosmic structures in Fig.~\ref{fig:tagging} in a $0.2 \; h^{-1} \rm Mpc$ slice for both the DM-only and the hydrodynamic simulations. We also show the regions with different tags in the 2 simulations. The change roughly traces the overdense regions, though some offset from the underlying density field is visible. 

\subsection{Statistics}
\label{subsec:statistics}
The effect of feedback on the power spectrum is quantified with the cross-power spectrum between the DM-only simulation and hydrodynamic simulation, 
\begin{equation}
    {P}_{\rm dm \; x \; hydro}(k) = \langle |\hat{\rho}_{\rm dm}(\mathbf{k})\hat{\rho}_{\rm hydro}(\mathbf{k})| \rangle .
\end{equation}
In order to separately examine the contribution of each type of structure, we create modified ``hydro'' grids, where only one specific structure (halo, filament, wall, or void) is from the hydrodynamic simulation, and the rest of the voxels are identical to the DM-only simulation. We compute the cross-power spectra between the DM grid and each of the 4 modified ``hydro'' grids using \textsc{nbodykit}\cite{nbodykit_Hand_2018}. 

In addition, we computed the PDF of the density field. While the power spectrum and its real space counterpart, the two-point correlation function, have been the default in cosmological analysis, in recent years, studies have found that there is rich, additional information in the nonlinear regime beyond the second order. The PDF is a simple statistic that contains non-Gaussian information and has been studied intensively for both 3D fields and 2D projected fields~\cite{pdf_Bernardeau2000,pdf_Liu2016,pdf_liumadhavacheril2019,pdf_Patton2017,pdf_Thiele2019,pdf_Friedrich2020,pdf_Thiele2020,pdf_Uhlemann_2020,pdf_Boyle2021,pdf_Barthelemy2021,pdf_Repp2021,pdf_Bernardeau2022,pdf_Uhlemann_2022}. We compute the PDF of the normalized density $\rho/\bar{\rho}$ in 100 logarithmically spaced bins spanning the range $\rho/\bar{\rho}\in[10^{-3}, 10^{3}]$. We again examine the change in PDF due to baryons for individual structures.

\section{Results}
\begin{figure*}[ht]
    \centering
    \includegraphics[width=2\columnwidth]{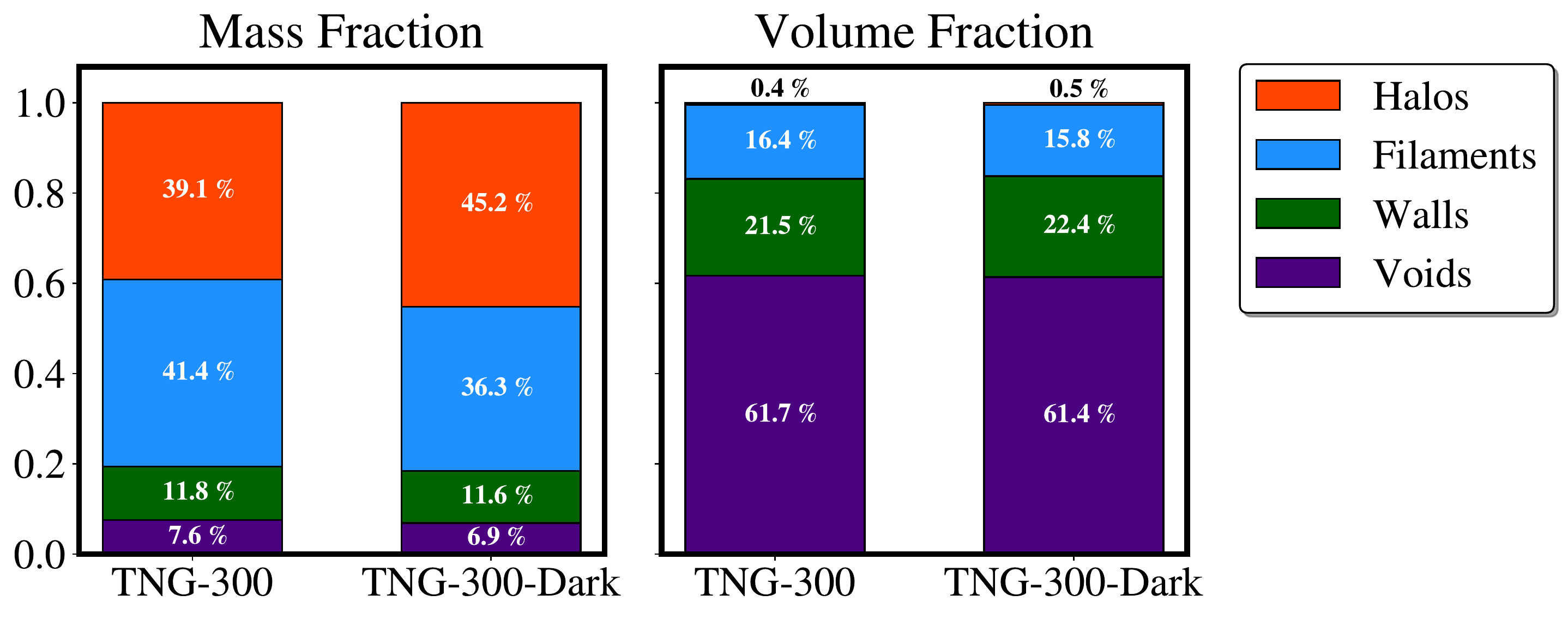}
    \caption{The mass (left) and volume (right) fraction each cosmic structure occupies, in TNG-300 and TNG-300-Dark simulations. Here we use $M_{\rm halo} > 10^{11} M_\odot$ in our structure tagging scheme (Sec.~\ref{subsec: nexus}). }
    \label{fig:total pie chart}
\end{figure*}

\label{sec:results}
Here we present the results of our analysis on the effects of baryonic feedback on the cosmic web, quantified as changes in mass and volume fractions of the structures, the power spectrum, and the PDF.

\subsection{Mass and Volume Fractions}
We show the mass and volume fractions of the cosmic structures in Fig.~\ref{fig:total pie chart}, for both the DM-only and hydrodynamic simulations. We found that 80\% of the mass in the universe is in halos and filaments, and the rest in walls and voids. In contrast, the space is mostly occupied by voids, with most of the remaining volume in walls, and filaments, leaving only half of a percent or less to halos. Here, we define halos with minimal mass $M_{\rm halo} > 10^{11} M_\odot$. Our findings are in general agreement with previous studies~\cite{Cosmic_Web_calvo, Cosmic_Web_Cautun, Cosmic_Web_Romero, Cosmic_Web_Shandarin, Cosmic_Web_Sousbie}. 

When we examine the changes in mass fractions, we found that halos lost more than 10\% of the mass due to baryons. Most of the lost mass is transferred to filaments and a small portion to walls and voids. This is expected as feedback expels matter from halos to the surrounding environment, which is most likely filaments. In addition, due to the reduced mass, some halos fall below our mass cut and hence lose their ``halo'' tag. When we examine the change in the volume fractions, baryons have negligible effect --- less than 1\% of the total simulation volume for four cosmic structures.  This indicates that the mass transfers we see do not significantly shift the boundaries of structures.  

\subsection{Power Spectrum}
\label{subsec:pk}

\begin{figure}[ht]
    \centering
    \includegraphics[width=\columnwidth]{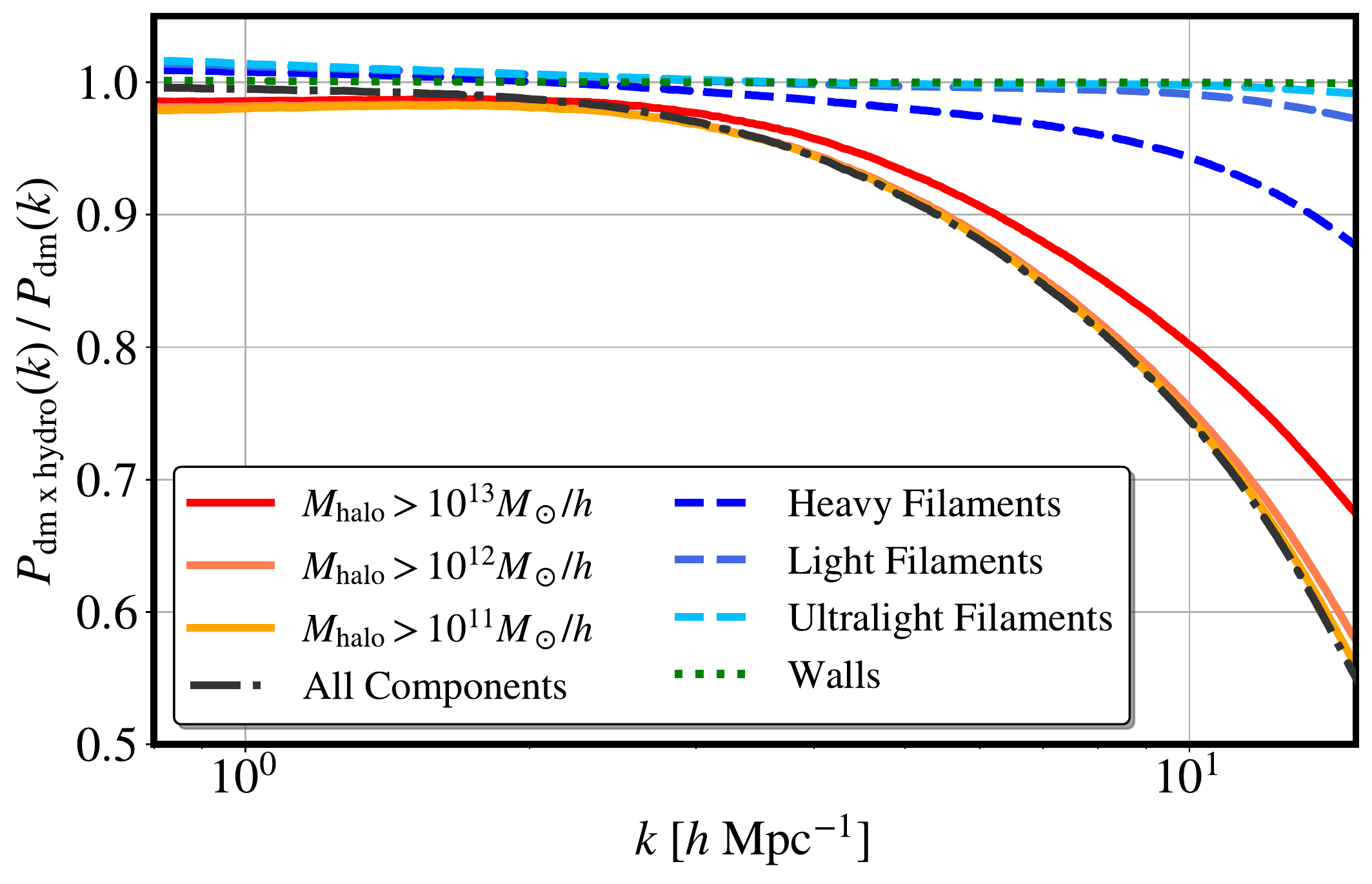}
    \caption{The ratio of the cross-power spectra of DM-only and hydrodynamic simulations with respect to the DM-only auto-power spectrum. Each curve includes one cosmic structure from the hydro run, while keeping the remaining voxels identical to the DM-only run: halos with $M_{\rm min}$=$[10^{11}, 10^{12}, 10^{13}]~M_\odot/h$ (red, orange, yellow solid lines), heavy, light, and ultralight filaments corresponding to the halo mass cuts above (dark, medium, light blue dashed lines), walls ({green} dotted lines). The total effect including all components is shown as the dark grey dash-dotted line.}
    \label{fig:isolated pk}
\end{figure}

We show the ratios of the cross-power spectra with respect to the DM-only power spectrum in Figure~\ref{fig:isolated pk}. We show results for 3 different halo mass cuts $M_{\rm min}$=$[10^{11}, 10^{12}, 10^{13}]~M_\odot/h$. While $10^{13}~M_\odot/h$ is where halo masses can be reasonably measured using weak lensing~\cite{Huang2020}, $10^{11}~M_\odot/h$ corresponds to the usual minimal halo mass in cosmological N-body simulations used by survey cosmology~\cite{sims_McCarthy2017,sims_Liu2018,sims_Harnois2018,sims_VN2021,sims_Angulo2021,sims_Maksimova2021,sims_Kacprzak2022}. The halo mass cut would affect other structures, as our algorithm identifies structures hierarchically --- first halos, second filaments, third walls, finally voids --- which means particles residing in halos with masses lower than $M_{\rm min}$ would be considered different structures, most likely filaments. Therefore, we also show three filament curves corresponding to the three different halos mass cuts. 

We find that baryonic effects suppress matter clustering in halos by 20\% and that in (heavy) filaments by 5\%, for $M_{\rm min}$=$10^{13}~M_\odot/h$ at $k=10 h^{-1} \mathrm{Mpc}$. When we lower the mass cut to $10^{12}~M_\odot/h$, the suppression on the power spectrum due to baryons is mostly accounted for by halos. Baryons have negligible effects on the wall or void (the latter not shown in the figure for legibility). 

\subsection{PDF}
\label{subsec:pdf}
\begin{figure}[ht]
    \centering
    \includegraphics[width=\columnwidth]{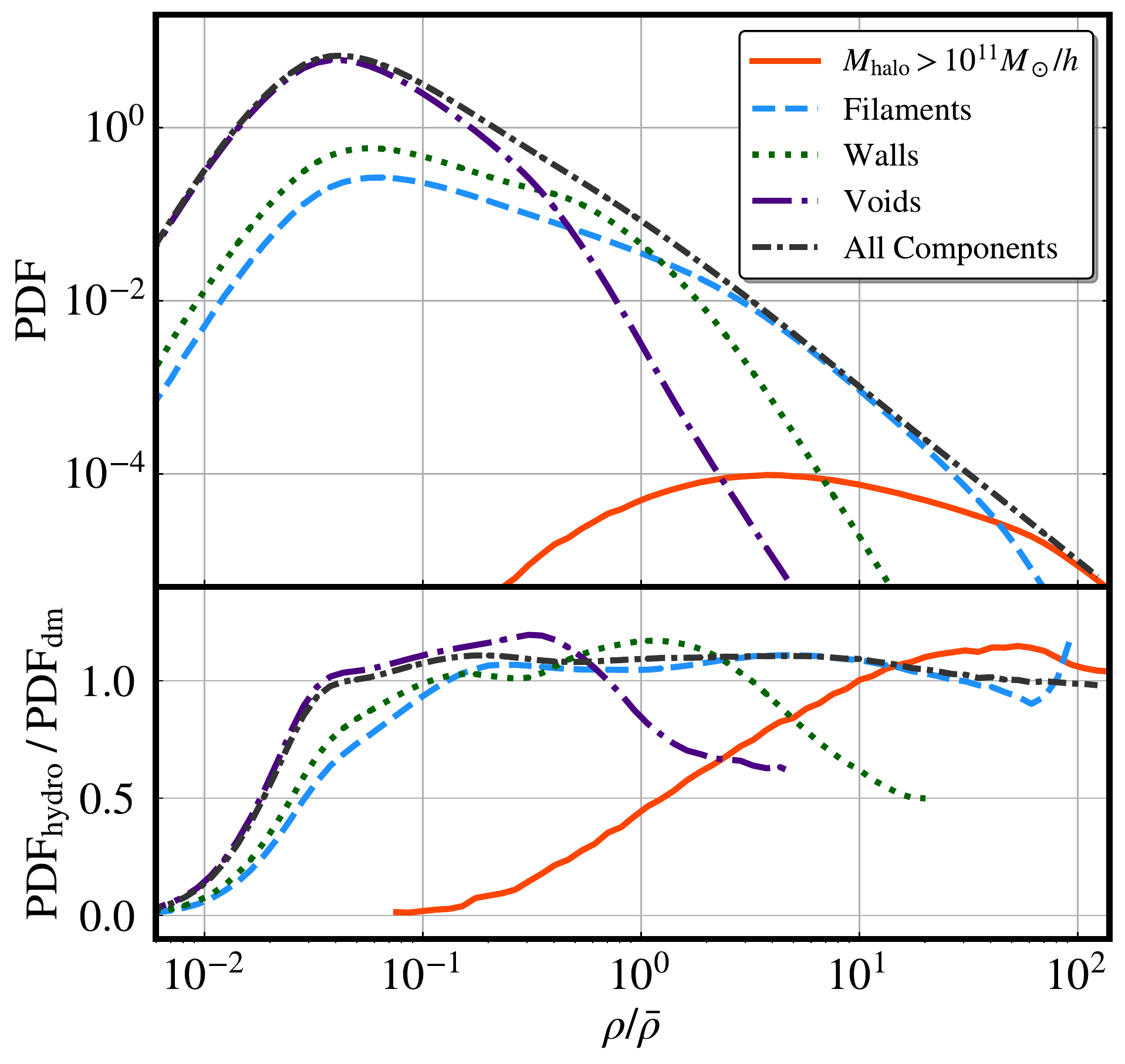}
    \caption{The PDF of the matter density in TNG-300 (top) and the ratios between the hydrodynamic and DM-only simulations (bottom) for all components (dark grey dash-dotted line) individual cosmic structures: halos with $M_{\rm min}$=$10^{11} M_\odot/h$ (red solid line), filaments (blue dashed line), walls (green dotted line), and voids (purple dash-dotted line).}
    \label{fig:PDF isolated}
\end{figure}

We show results of PDF of the density field in Figure \ref{fig:PDF isolated}. 
Unlike the power spectrum, whose signal is dominated by overdense regions, the PDF weighs volume equally and disregards the matter content inside. For halos, we see a boost for high-density regions and suppression of low-density regions. This could be due to the redistribution of matter within halos, where the inner regions become denser due to radiative cooling and outer regions less dense due to stellar and AGN feedback. 

It is interesting to see that in the $\rho/\bar{\rho}\in [10^{-1},10^1]$ regions, while the halo PDF is suppressed by baryonic feedback, the filament PDF is boosted by $\approx10\%$. Filaments are likely the structures immediately adjacent to halos and hence receive the injected mass from baryonic feedback. However, denser wall or void regions are suppressed, followed by a boost when moving towards less dense regions. At the emptiest regions of the universe $\rho/\bar{\rho}<10^{-2}$, baryons universally suppress the densities of all structures. 

Despite seeing negligible baryonic effects for walls and voids in the power spectrum, we find nearly 100\% suppression of the emptiest regions and 10\%-level effects (boost or suppression) in the remaining regions of filaments, walls, and voids. Our findings show that future cosmological surveys using the PDF will need to take into consideration diffused structures beyond halos. 

\section{Conclusion}
\label{sec:conclusion}
In this work, we study the effect of baryons on the cosmic web --- halos, filaments, walls, and voids --- by applying a modified version of the NEXUS algorithm on the IllustrisTNG simulations (Fig.~\ref{fig:tagging}). Our main findings are:
\begin{itemize}
    \item Halos lose more than 10\% of their mass due to baryons. Most of the lost mass is transferred to filaments and a small portion to walls and voids. However, the mass transfer does not significantly shift the boundaries of structures (Fig.~\ref{fig:total pie chart}).
    \item Most suppression in the power spectrum can be accounted for by $M\ge10^{12}~M_\odot/h$ halos. However, if only $M\ge 10^{13}~M_\odot/h$ halos are modeled, one would underestimate the total suppression of the matter power spectrum to be 20\% instead of 25\% at $k=10 h^{-1} \mathrm{Mpc}$. Baryons have negligible effects on the power spectrum of walls or voids (Fig.~\ref{fig:isolated pk}). 
    \item When examining the PDF of the density field, we find nearly 100\% suppression of the emptiest regions and 10\%-level effects (boost or suppression) in the remaining regions of filaments, walls, and voids, indicating that cosmological analysis with PDFs will need to take into consideration diffused structures beyond halos (Fig.~\ref{fig:PDF isolated}). 
\end{itemize}

In our work, we only applied our method to the IllustrisTNG simulations, so it would be beneficial to validate against other hydrodynamic simulations with different subgrid models in the future. It would also be interesting to study the redshift evolution of the baryonic effects on the cosmic web, which may be useful to help break the degeneracy between baryonic effects and cosmological parameters. In summary, our work demonstrates the importance of modeling the baryonic effects in the whole cosmic web, not just halos, for cosmological analysis beyond two-point statistics or field-based inferences. 

\begin{acknowledgments}
This research used resources of the National Energy Research Scientific Computing Center (NERSC), a U.S. Department of Energy Office of Science User Facility located at Lawrence Berkeley National Laboratory, operated under Contract No. DE-AC02-05CH11231. 
\end{acknowledgments}

\bibliographystyle{physrev}
\bibliography{paper}
\end{document}